\documentclass[twocolumn,showpacs,preprintnumbers,amsmath,amssymb]{revtex4}

\usepackage{graphicx}
\usepackage{dcolumn}
\usepackage{bm}
\usepackage{epsfig}
\usepackage{color}

\begin{document}


\title{Eshelby ensemble of highly viscous flow out of equilibrium}

\author{U. Buchenau}
 \email{buchenau-juelich@t-online.de}
\affiliation{%
Forschungszentrum J\"ulich GmbH, J\"ulich Centre for Neutron Science (JCNS-1) and Institute for Complex Systems (ICS-1),  52425 J\"ulich, GERMANY
}%

\date{April 28, 2019}

\begin{abstract}
The recent description of the highly viscous flow in terms of irreversible structural Eshelby rearrangements is extended to calculate the heat capacity of a glass former at a constant cooling rate through the glass transition. The result is compared to measured data from the literature, showing that the explanation works both for polymers and other glass formers. 
\end{abstract}

\maketitle

It is a generally acknowledged fact that the undercooled liquid falls out of the thermal equilibrium at the glass transition temperature $T_g$, keeping its atomic structure. The glass transition is kinetic; the transition temperature $T_g$ is lower at a smaller cooling rate \cite{stillinger,ediger,angell,schick1,cavagna,schmelzer}.

Below the glass temperature, the description of the frozen glass on its slow way back to equilibrium requires an additional parameter, the fictive temperature \cite{tool,nara,hodge} $T_f$, which is higher than the phonon temperature $T$ and characterizes the frozen state of the structural degrees of freedom. 

The main problem for our understanding of the glass transition is that there is no detailed physical picture of the structural relaxation. If one knows the physical mechanism of a transition, it is straightforward to treat non-equilibrium situations \cite{odagaki}. But this is not yet the case for the glass transition.

An appealing candidate for a convincing physical mechanism of the glass transition is the cooperative shear transformation of an atomic cluster in Fig. 1, first proposed as "cooperative shear zone" to explain the temperature dependence of the viscosity of metallic glass formers \cite{johnson}. The cooperative shear transformation leads to a change of the elastic shear misfit of the cluster with respect to the surrounding viscoelastic matrix, tractable in terms of the Eshelby theory \cite{eshelby}. 

\begin{figure}   
\hspace{-0cm} \vspace{0cm} \epsfig{file=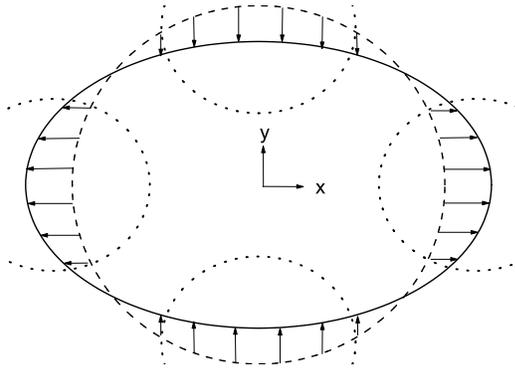,width=7 cm,angle=0} \vspace{0cm} \caption{Cooperative structural shear transformation of an Eshelby region embedded in the supercooled liquid.}
\end{figure}

Very recently, the distinction between reversible and irreversible processes has been introduced into the Eshelby shear transformation picture, with the irreversible processes responsible for the viscous flow, and the reversible ones responsible for the recoverable shear compliance $J_0$ and the short-time Kohlrausch time dependence \cite{asyth,asyth1}.

The irreversible Eshelby mechanism of the highly viscous flow \cite{asyth} simplifies the real liquid to one composed of domains which are large enough to undergo irreversible structural Eshelby shear transitions. The size of these atomic clusters seems to be about forty atoms in simple liquids and about twenty monomers in polymers (see Supplemental Material). The Eshelby domains are characterized exclusively by the elastic shear misfit energy $e^2$ (in units of $k_BT$). All other energy contributions are neglected. The size is measured by the number $N$ of atoms, molecules or monomers within the region. The terminal relaxation time $\tau_c$ of the liquid, which marks the transition from reversible to irreversible Eshelby transitions, is supposed to correspond to the size $N_c$.

In the simplest form of the Eshelby theory \cite{eshelby}, the shear misfit energy $e^2$ is given by
\begin{equation}
e^2=\frac{GN_cV_a\epsilon^2}{4k_BT},
\end{equation}
where $G$ is the short time shear modulus, $V_a$ is the volume of a single atom, molecule or monomer and $\epsilon$ is the shear angle misfit between the Eshelby domain and the viscoelastic surroundings.

Since there are five independent possible shears for a given region, the space of possible $e$-states is five-dimensional. A constant density of stable structures in this five-dimensional $e$-space is assumed, together with an energy barrier between them which depends only on the size \cite{adam}.

With these assumptions, the average shear misfit energy is $\overline{e^2}=5/2$ in the normalized distribution
\begin{equation}\label{pe}
	p(e)=\frac{1}{\pi^{5/2}}e^4\exp(-e^2).
\end{equation}

Regions with a strong elastic shear misfit have a shorter lifetime than the weakly strained ones, because they decompose predominantly by down-jumps in energy, which have a higher rate than up-jumps (the jump rate from the state $e_0$ to $e$ has the average factor $\exp((e_0^2-e^2)/2)$ from the elastic misfit difference). The detailed calculation \cite{asyth} yields the lifetime $\tau_e$ for the state with the energy $e^2$  
\begin{equation}\label{taue}
\tau_e=4\sqrt{2}\tau_c\exp(-e^2/2)
\end{equation}
and the viscosity $\eta$ from the irreversible shear fluctuations
\begin{equation}\label{etafluct}
	\frac{\eta}{G}=\frac{1}{8}\tau_c.
\end{equation}

Translating $p(e)$ into a distribution $p(\ln{\tau})$ for the lifetimes in thermal equilibrium, one gets
\begin{equation}\label{pt}
p(\ln{\tau})=\frac{\tau^2}{3\sqrt{2\pi}\tau_c^2}\left(\ln{\frac{4\sqrt{2}\tau_c}{\tau}}\right)^{3/2}.
\end{equation}
In time, the resulting relaxation function is close to a Kohlrausch function $\exp(-(t/\tau)^\beta)$ with $\tau=1.783\tau_c$ and $\beta=0.824$.

The strongest evidence for the validity of this irreversible Eshelby mechanism for the highly viscous flow comes from the comparison of shear mechanical and calorimetric data. Measurements of $G(\omega)$ provide $G$ and $\eta$, which allows to calculate $\tau_c$ from eq. (\ref{etafluct}) with reasonable accuracy. With this $\tau_c$, one finds a very satisfactory description of calorimetric TMDSC (Temperature Modulated Differential Scanning Calorimetry) thermal equilibrium data at the same temperature, using the relaxation time distribution of eq. (\ref{pt}). It is a bit surprising that no trace of the reversible processes is seen in the calorimetric data, but it seems to be an experimental fact.

This was first shown \cite{asyth} for PPE, a vacuum pump oil consisting of five connected phenylene rings and later \cite{asyth1} for a metallic glass, for glycerol and for propylene glycol. The present paper will add more examples.

But the main question of the present paper is whether the irreversible Eshelby mechanism is also able to describe the fall out of equilibrium at the glass transition. 

The best measurements to test this are the pioneering calorimetric data of Hensel and Schick \cite{hensel}. Hensel and Schick did a systematic comparison of data in thermal equilibrium with out-of-equilibrium data taken at a constant cooling rate through the glass transition. This was done for the two polymers polystyrene and polyether ketone, the silicate glass 2SiO$_2$-NaO, a commercial window glass DGG-STG1, and the ionic glass former CKN, a mixture of potassium and calcium nitrate.  

In their work \cite{hensel}, the equilibrium data are TMDSC scans, measuring the real and imaginary parts $c_p'$ and $c_p''$, respectively, at a given frequency $\omega$ as a function of temperature. The non-equilibrium data are temperature scans of $c_p(T,q)$ at a given cooling rate $q$.

The equilibrium data of Hensel and Schick are again well described in terms of the irreversible Eshelby mechanism.
Integrating over the distribution of eq. (\ref{pt}), one finds the equilibrium $c_p''(\omega)$-peak of a TMDSC scan at constant frequency at the temperature $T_\alpha$ with
\begin{equation}\label{oma}
	\omega\tau_c(T_\alpha)=0.533.
\end{equation}

\begin{table}[htbp]
	\centering
		\begin{tabular}{|c|c|c|c|c|}
\hline
substance   & $B$          &   $T_{VF}$          &   $\Delta T_{exp}$   &  $\Delta T_{calc}$\\
\hline   
            &   K          &     K               &        K             &         K         \\
\hline
polystyrene & 1837         & 319                 &         2.6$\pm$0.2  &          2.75     \\
CKN         & 1594         & 291                 &         3.0$\pm$0.3  &          2.46     \\
2SiO$_2$-NaO& 8826         & 483                 &        10.5$\pm$1.5  &         13.3      \\
DGG-STG1    &10161         & 532                 &        14.0$\pm$1.5  &         15.4      \\
polyether ketone&1627      & 378                 &         2.0$\pm$0.2  &          2.46     \\
\hline
		\end{tabular}
	\caption{Vogel-Fulcher parameters, mixing coefficient $b$ and a comparison of measured and calculated widths $\Delta T$ of constant-frequency scans for the five substances of Hensel and Schick \cite{hensel}.}
	\label{tab:C}
\end{table}

One begins by the determination of the equilibrium-$\tau_c$ from the TDMSC scans, using eq. (\ref{oma}). The equilibrium values found in this way are fitted by a Vogel-Fulcher law
\begin{equation}\label{vf}
	\ln{\tau_c}=\frac{B}{T-T_{VF}}-13\ln{10}
\end{equation}
corresponding to the Arrhenius barrier
\begin{equation}\label{vct}
   V_c(T)=Bk_BT/(T-T_{VF}).	
\end{equation}

Having the two Vogel-Fulcher parameters $B$ and $T_{VF}$, one knows $\tau_c(T)$ at all temperatures and can calculate $c_p'(\omega)$ and $c_p''(\omega)$ as a function of temperature for a given frequency $\omega$. The calculated data are found to be well described by a gaussian with exponent $(T-T_\alpha(\omega))^2/2\Delta T^2$, the fit function used in the experiment \cite{hensel}.

Table I lists the two Vogel-Fulcher parameters and compares calculated (frequency 0.1047 rad/s) and measured \cite{hensel} widths $\Delta T$.

To describe the non-equilibrium data, one has to consider the Eshelby ensemble and its average fictive temperature $T_f$ for a given cooling rate $q=\partial T/\partial t$ in the neighborhood of the temperature $T_g(q)$, where the excess heat capacity $\Delta c_p$ of the undercooled liquid over the phonon heat capacity of the glass reaches half of its thermal equilibrium value. At this temperature, $\dot{T_f}=q/2$.

The time development of the average fictive temperature follows from the differential equation
\begin{equation} \label{diff}
	\dot{T_f}=\frac{T-T_f}{\tau_c(T_f,T)}
\end{equation}
because $\tau_c$ was defined in reference \cite{asyth} as the inverse of the average decay rate.

Close to equilibrium, for $T$ closely below $T_f$, the relaxation time $\tau_c(T_f,T)$ of the Eshelby ensemble is given by the Arrhenius relation
\begin{equation}\label{tctft}
\tau_c(T_f,T)=\tau_0\exp(V_c(T_f)/k_BT),	
\end{equation}
with the energy barrier $V_c(T_f)$, which can be determined from eq. (\ref{vct}), and $\tau_0=10^{-13}$ s. 

This simple scheme is good enough to calculate $T_g(q)$ for cooling scans, but fails when $T$ leaves the neighborhood of $T_f$. Remember that the Eshelby domain converts from one equilibrated at $T_f$, containing $N_c(T_f)$ particles, to one at equilibrium at $T$, containing $N_c(T)$ particles. At lower temperatures, this means that the domain does not only have to rearrange, but to increase as well. A detailed treatment \cite{ast} of heating data in polystyrene \cite{tropin} shows that one can take this influence into account by postulating a mixing coefficient $b$ of the barriers at the two temperatures
\begin{equation}\label{vctf}
	V_c(T_f,T)=bV_c(T_f)+(1-b)V_c(T),
\end{equation}
with $b=2/3$. Eq. (\ref{vctf}) returns to eq. (\ref{vct}) when the two temperatures are close together.

\begin{figure}   
\hspace{-0cm} \vspace{0cm} \epsfig{file=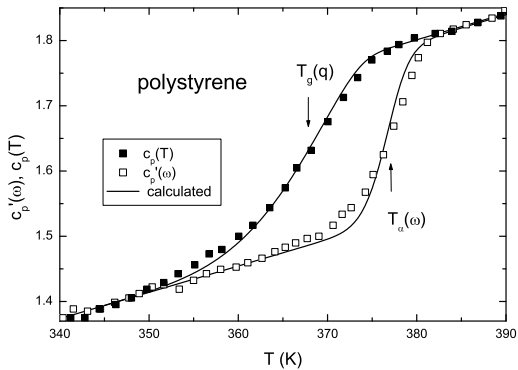,width=7 cm,angle=0} \vspace{0cm} \caption{Out-of-equilibrium data ($T_g(q)$ at $q=0.5$ K/min) and equilibrium data ($T_\alpha(\omega)$ at $\omega=0.1047$ rad/s) from the same polystyrene sample \cite{hensel}. The figure demonstrates that the irreversible Eshelby mechanism \cite{asyth} is able to reproduce the cooling curve with eq. (\ref{vctf}) and $T_{VF}$=217 K.}
\end{figure}

Replacing $V_c(T_f)$ by this $V_c(T_f,T)$ in eq. (\ref{tctft}), one gets the $\tau_c(T_f,T)$ which one has to insert into eq. (\ref{diff}) to calculate the time dependence of the average fictive temperature at a constant cooling rate $q=\dot{T}$, beginning at high temperature where the difference between $T_f$ and $T$ is negligible. The differential equation is solved numerically for small temperature steps, integrating the exponential decay over the corresponding time step with the average $\tau_c$ between the two temperatures.

Fig. 2 compares measured \cite{hensel} and calculated curves, both in equilibrium and out of equilibrium, for polystyrene, showing that the form of the measured curves is well described by the irreversible Eshelby scheme, both in equilibrium and out of equilibrium. Out of equilibrium, it is necessary to adapt $T_{VF}$ to the curve; this is also seen in Fig. 3, where the point at $q=0.5$ K/min lies two degrees K to the left of the calculated curve.

\begin{figure}   
\hspace{-0cm} \vspace{0cm} \epsfig{file=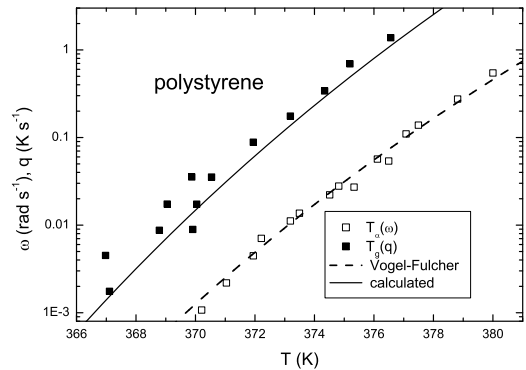,width=7 cm,angle=0} \vspace{0cm} \caption{$T_g(q)$ and $T_\alpha(\omega)$ for different cooling rates and frequencies in polystyrene \cite{hensel} calculated with the parameters in Table I.}
\end{figure}

\begin{figure}   
\hspace{-0cm} \vspace{0cm} \epsfig{file=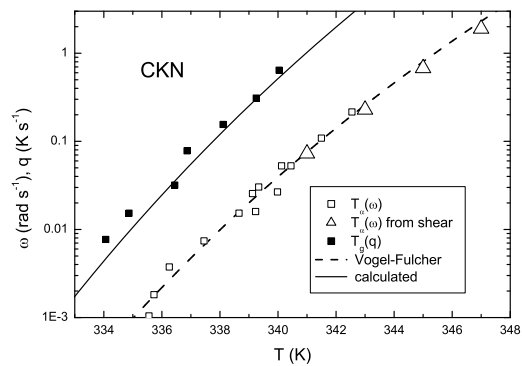,width=7 cm,angle=0} \vspace{0cm} \caption{$T_g(q)$ and $T_\alpha(\omega)$ for different cooling rates and frequencies in the ionic glass former CKN \cite{hensel} (parameters see Table I). Note the good agreement of the calorimetric $T_\alpha(\omega)$ with those calculated via equs. (\ref{etafluct}) and (\ref{oma}) from a recent dynamical shear measurement \cite{bierwirth}.}
\end{figure}

Fig. 4 shows the result for CKN, including a comparison to four $T_\alpha(\omega)$-values calculated from a recent dynamical shear measurement \cite{bierwirth}, which provided $\tau_c$ at four temperatures via eq. (\ref{etafluct}). Similar agreement is found for the two silicate glass formers and polyether ketone, as shown in the Supplemental Material. In all these calculations, the condition
\begin{equation}
	\dot{\tau_c}=1.54
\end{equation}
was found to be fulfilled at $T_g(q)$. This holds also if one varies the mixing coefficient $b$ in eq. (\ref{vctf}).

The agreement between measured and calculated $T_g(q)$ in the five very different glass formers is achieved with only one assumption: the validity of the Vogel-Fulcher law, leading to eq. (\ref{vct}), proved in the past for countless examples. Therefore the agreement is convincing evidence for the validity of the irreversible Eshelby mechanism of the highly viscous flow.

A second important consequence, already perceived in preceding work \cite{ngai,beiner,schweizer1,schweizer,sokolov1,sokolov}, concerns the polymer case. Obviously, the calorimetric data see only the irreversible part of the segmental relaxation and not the relaxation of the whole polymer. This irreversible segmental relaxation is well described by the irreversible Eshelby rearrangement of a small region, in a polymer as well as in any other glass former.

After the segmental relaxation, the long polymer chain is not yet relaxed, but relaxes further in the so-called Rouse modes, leading to a polymer viscosity orders of magnitude larger than the one calculated from eq. (\ref{etafluct}) with the parameters in Table I. 

The analysis of polystyrene mechanical data separating segmental relaxation and Rouse modes \cite{ngai} arrives at $GJ_0\approx 3.5$ for the segmental relaxation. Indeed, the condition $GJ(\omega)=3.5$ for polystyrene \cite{beiner} lies close to $\omega\tau_c=1$, with $\tau_c$ calculated from the Vogel-Fulcher parameters in Table I (see Supplemental Material).

To conclude, the irreversible Eshelby mechanism seems to be the adequate description of the highly viscous flow. It provides a quantitative description of calorimetric data, both in and out of thermal equilibrium. For polymers, the results strengthen the conclusion that the segmental relaxation has its own temperature dependence and is due to the same elementary process as in all other glass formers.

Thanks are due to Alexei Sokolov for pointing out the polystyrene problem and helpful discussions, to Christoph Schick for sending his papers, to Peter Bierwirth, Roland B\"ohmer and Catalin Gainaru for communicating their shear data prior to publication, and to Andreas Wischnewski for an illuminating discussion on reversible and irreversible recoverable compliance.


\begin{thebibliography}{99}
\bibitem{stillinger} F. H. Stillinger, Science {\bf 267}, 1935 (1995)
\bibitem{ediger} M. D. Ediger, C. A. Angell, and S. R. Nagel, J. Phys. Chem. {\bf 100}, 13200 (1996) 
\bibitem{angell} C. A. Angell, K. L. Ngai, G. B. McKenna, P. F. McMillan, and S. W. Martin, J. Appl. Phys. {\bf 88}, 3113 (2000)
\bibitem{schick1} C. Schick, {\it Temperature modulated differential scanning calorimetry (TMDSC) — basics and applications to polymers}, in: P.K. Gallagher, S.Z.D. Cheng (Eds.), Handb. Therm. Anal. Calorim. Appl. to Polym. Plast, Elsevalsoier, Amsterdam, 2002, pp. 713–810
\bibitem{cavagna} A. Cavagna, Phys. Rep. {\bf 476}, 51 (2009)
\bibitem{schmelzer} J. W. P. Schmelzer, I. S. Gutzow, O. V. Mazurin, A. I. Priven, S. V. Todorova, and B. P. Petroff, {\it Glasses and the Glass Transition}, Wiley-VCH Verlag, Weinheim, Germany, 2011
\bibitem{tool} A. Q. Tool, J. Am. Ceram. Soc. {\bf 29}, 240 (1946)
\bibitem{nara} O. S. Narayanaswamy, J. Am. Ceram. Soc. {\bf 54}, 491  (1971)
\bibitem{hodge} I. M. Hodge, J. Non-Cryst. Solids {\bf 169}, 211 (1994)
\bibitem{odagaki} T. Tao, A. Yoshimori, and T. Odagaki, Phys. Rev. E {\bf 64}, 046112 (2001)
\bibitem{johnson} M. D. Demetriou, J. S. Harmon, M. Tao, G. Duan, K. Samwer, and W. L. Johnson, Phys. Rev. Lett. {\bf 97}, 065502 (2006)
\bibitem{eshelby} J. D. Eshelby, Proc. Roy. Soc. {\bf A241}, 376 (1957)
\bibitem{asyth} U. Buchenau, J. Chem. Phys. {\bf 148}, 064502 (2018)
\bibitem{asyth1} U. Buchenau, J. Chem. Phys. {\bf 149}, 044508 (2018)
\bibitem{adam} G. Adam and J. H. Gibbs, J. Chem. Phys. {\bf 43}, 139 (1958)
\bibitem{hensel} A. Hensel and C. Schick, J. Non-Cryst. Solids {\bf 235-237}, 510 (1998)
\bibitem{ast} U. Buchenau, unpublished
\bibitem{tropin} T. V. Tropin, G. Schulz, J. W. P. Schmelzer, and C. Schick, J. Non-Cryst. Solids {\bf 409}, 63 (2015)
\bibitem{bierwirth} J. Beerwerth, S. P. Bierwirth, J. Adam, C. Gainaru, and R. B\"ohmer, private communication
\bibitem{ngai} K. L. Ngai, D. J. Plazek, and I. Echeverria, Macromolecules {\bf 29}, 7937 (1996)
\bibitem{beiner} M. Beiner, S. Reissig, K. Schr\"oter, and E.-J. Donth, Rheol. Acta {\bf 36}, 187 (1997)
\bibitem{schweizer1} S. Mirigian and K. S. Schweizer, J. Chem. Phys. {\bf 140}, 194507 (2014)
\bibitem{schweizer} S. Mirigian and K. S. Schweizer, Macromolecules {\bf 48}, 1901 (2015)
\bibitem{sokolov1} C. Dalle-Ferrier, A. Kisliuk, L. Hong, G. Carini, Jr., G. Carini, G. D' Angelo, C. Alba-Simionesco, V. N. Novikov, and A. P. Sokolov, J. Chem. Phys. {\bf 145}, 154901 (2016)
\bibitem{sokolov} A. L. Agapov, V. N. Novikov, T. Hong, F, Fan, and A. P. Sokolov, Macromolecules {\bf 51}, 4874 (2018)
\end{thebibliography}
\end{document}